\documentclass{svproc}
\usepackage{url}
\usepackage{wrapfig}
\usepackage{graphicx}

\begin{document}
\mainmatter              
\title{ALICE Inner Tracking System Upgrade: construction and commissioning}
\titlerunning{ALICE Inner Tracking System Upgrade: construction and commissioning}  

\author{D. Colella for the ALICE Collaboration}
\authorrunning{D. Colella} 

\institute{Istituto Nazionale di Fisica Nucleare, \\
via E. Orabona 4 \\
Bari, 70125, Italy\\
\email{domenico.colella@cern.ch}, \\
}

\maketitle              

\begin{abstract}
ALICE (A Large Ion Collider Experiment) is the CERN LHC experiment optimized for the study of the strongly
interacting matter produced in heavy-ion collisions and devoted to the characterization of the Quark-Gluon
Plasma. To achieve the physics program for LHC Run 3 and 4, ALICE is undergoing a major upgrade of the experimental
apparatus during the ongoing second long LHC shutdown.

A key element of the ALICE upgrade is the substitution of the Inner Tracking System (ITS) with a completely new
silicon-based detector whose features will allow the reconstruction of rare physics channels
which could not be accessed before with the ITS layout used during LHC Run 1 and 2. 
The enabling technology for such performance boost is the
adoption of custom-designed MAPS (Monolithic Active Pixel Sensors) as detecting element.

In this proceedings, the status of the construction and commissioning of the ITS upgrade will be detailed.
The completion of the modules construction will be achieved soon and, in the meantime, 
the commissioning in laboratory is proceeding using the components already integrated in the detector.

\end{abstract}

\section{ALICE Inner Tracking upgrade}
The main objective of the ALICE experimental program for LHC Run 3 and 4 is a detailed exploration of the Quark-Gluon
Plasma properties via high precision measurements of rare probes in \mbox{pp}, \mbox{p--Pb} and \mbox{Pb--Pb}
collisions \cite{ALICEupLoI}. In order to be able to measure a large sample of short lived systems such as
heavy-flavor hadrons, quarkonia, and low mass dileptons over a wide range of transverse momenta, it is necessary to
enhance the tracking and readout rate capabilities with respect to the ones of the detector that worked during
the LHC Run 1 and 2. During the LHC Long Shutdown 2 in 2019-2020, several sub-detectors will be upgraded and the
Inner Tracking System (ITS) will be replaced.

\begin{figure}[t]
\centering
\includegraphics[width=30pc,clip]{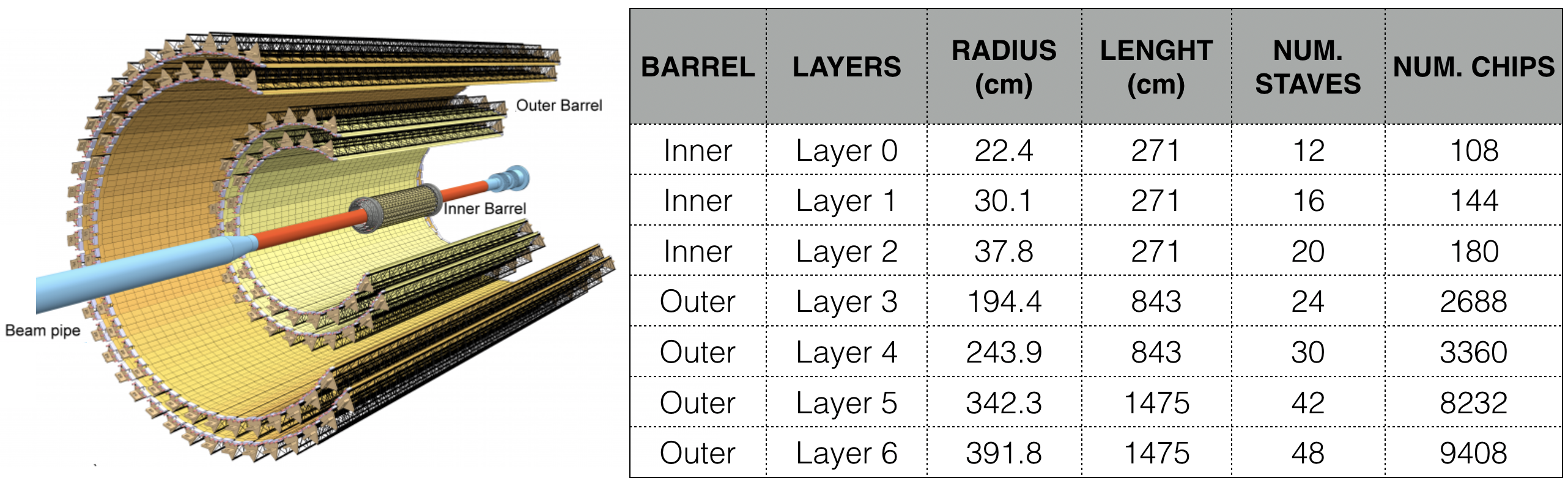}
\caption{(Left) Schematic layout of the upgraded ITS. (Right) Summary table containing the detector layers geometric 
characteristics.}
\label{fig-1}
\end{figure}

The key improvements of the ITS upgrade are \cite{ITSupTDR}: a completely new layout envisaging seven cylindrical layers
(three innermost layers referred as Inner Barrel, IB, and four outermost layers referred as Outer Barrel, OB) all equipped
with ALPIDE Monolithic Active Pixel Sensors (MAPS) chips of about 30x30 $\mu$m$^{2}$ pixel size, a large reduction of 
the material
budget ($\sim$0.3\% X$_{0}$ for the IB and $\sim$1\% X$_{0}$ for the OB), and a reduced distance of the innermost layer to the
interaction point. These design characteristics will allow a strong improvement of the
tracking efficiency at low transverse momenta as well as the impact parameter
resolution \cite{ITSupPaolo}. 
A schematic view of the detector layout and a table reporting geometrical details for each layer
are shown in Figure \ref{fig-1}.

The very low value for the material budget has been obtained adopting monolithic silicon pixel chips, integrating the 
sensor and the readout electronics functionalities in the same substrate, thinned to 50 $\mu$m in the IB (100 $\mu$m in 
the OB). The pixel chips are mounted on low-mass polyimide Flexible Printed Circuits 
(FPC, for powering and data stream), glued on a carbon fiber structure (space frame), equipped with water cooling pipes 
integrated in the structure. 

The chips (9 for the IB and 14 for
the OB) glued and wire-bonded to the FPC constitute the Hybrid Integrated Circuit (HIC). The HIC(s) (1 for the IB, 8 or
14 for the two innermost and the two outermost layers of the OB respectively) glued to the space frame constitute the Stave.
More details on the ALPIDE chip characteristics and R\&D can be found in \cite{AlpideGAR}.

\section{Component production status}
The electrical and functional tests of the ALPIDE chips have been performed at CERN (CH) for the 50 $\mu$m thinned
chips and at Yonsei and Pusan/Inha (KR) for the 100 $\mu$m thinned chips, between
September 2017 and May 2019. A total of almost 4$\times{10^{4}}$ chips have been tested with a detector-grade quality
yield of 63.7$\%$.

The full production of HICs and Staves for the IB has been carried out at CERN and concluded at mid 2019. A total of
95 Staves, enough to build two copies of the three inner barrel layers, have been assembled with a yield of 73$\%$.
The OB HICs have been produced in five sites (Bari (IT), Liverpool (UK), Pusan/Inha, Strasbourg (FR) and Wuhan(CN)), while 
the FPCs have been tested in Trieste (IT). It took 80 weeks to assembly the needed 2500 HICs; this quantity includes 
the HICs needed to cover the whole detector acceptance plus spares and assumes an overall production yield of 74$\%$ 
(convolution of 82$\%$ for the HIC and 90$\%$ for the Stave). 
The HIC production has been completed and a yield of 84$\%$ has been achieved.
The OB Staves have been assembled in five sites (Berkeley (US), Daresbury (UK), Frascati (IT), NIKHEF (NL) and Torino (IT)).
Enough Staves to cover the full OB acceptance have been assembled and qualified as detector grade. The Stave assembly
yield is close to 90$\%$ and production of few more spare Staves will be completed by October 2019.

Full readout logic is implemented in the ALPIDE chip that sends the digitized and zero-suppressed hit data to the
off-detector electronics. A total of 192 FPGA-based readout units \cite{ITSRUJo} (CERN, Bergen (NO), NIKHEF) 
control and monitor
the sensors and their power supply modules, receive the trigger and detector control information, and deliver the sensor data
to the counting room. A total of 142 boards (Berkeley), able to provide analog and digital 1.8 V supplies to each HIC plus a
negative voltage output for the reverse bias, are needed to power the full detector. Production and qualification of
the full set of readout and power boards are completed.

Detailed description of the mechanical support structure layout can be found in \cite{ITSupTDR}. All the components
have been produced and verified (Berkeley, CERN, Padova (IT)). A dry insertion test of a dummy version of
IB half-detector has been successfully carried out.
Cables placing in the supporting structures is ongoing and required smart solutions to fit all in the best way.

\section{Layer assembly and commissioning}
A large clean room has been built at CERN to allow the full detector assembly and the on-surface commissioning activities,
before the installation in the ALICE cavern in May 2020. Here the same backend system that will be used in the experiment
is available, including powering system, cooling system, full readout chain. Integration of the Staves in the layer structure
and connection to the services is done once the Staves are shipped to CERN, in parallel with the commissioning on the
already installed part of the system.

\begin{figure}[t]
\centering
\includegraphics[width=30pc,clip]{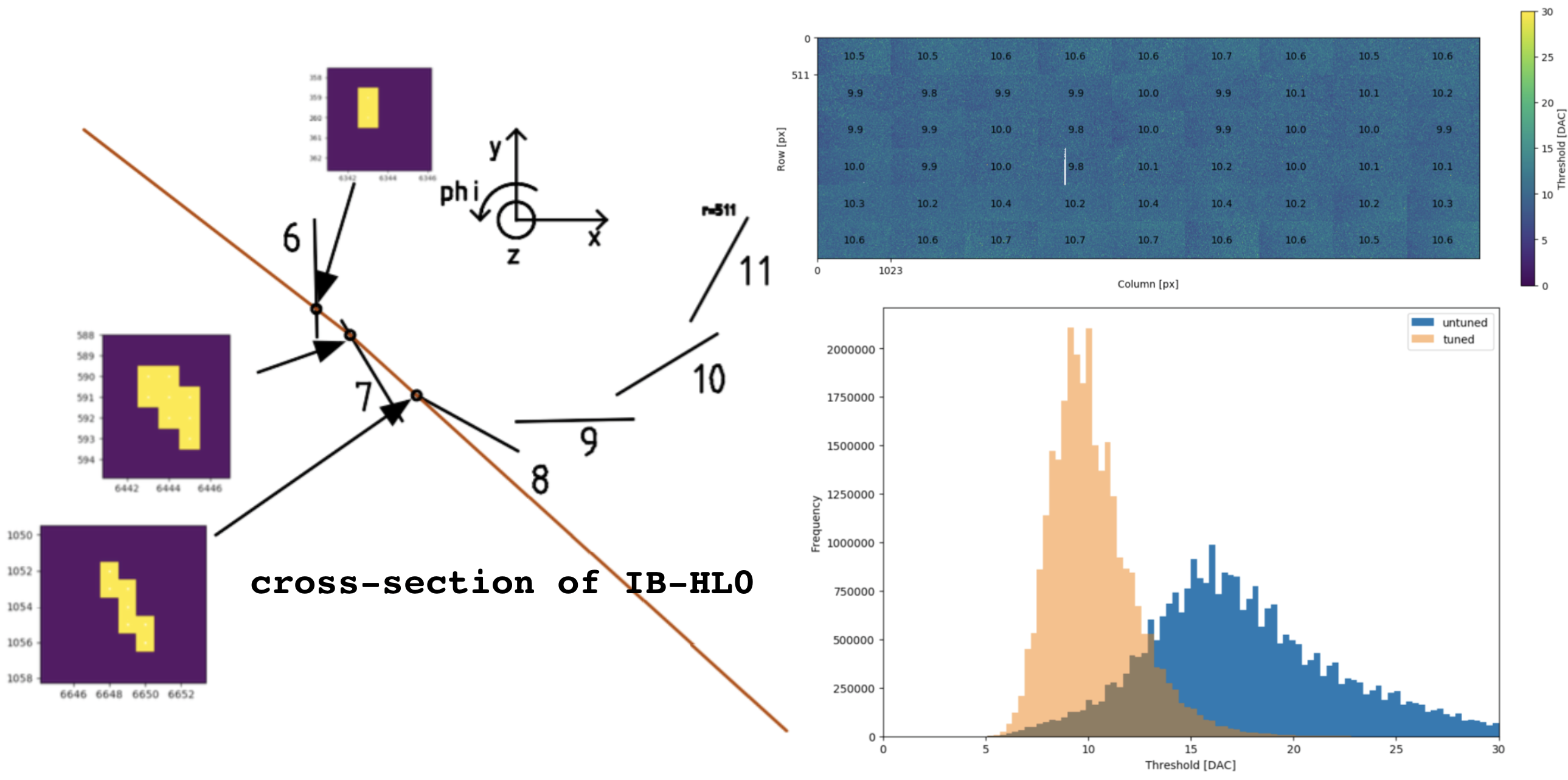}
\caption{(Left) First cosmic ray identified in the IB-HL0, including cluster shape of the track hit in the three crossed Staves.
         (Right) Map of the threshold value for all the pixels in the six Staves included in the IB-HL0 after tuning of the DAC parameters.}
\label{fig-2}
\end{figure}

Detector Control System (DCS) and Data Acquisition System (DAQ) softwares are available and running on machines housed
in a control room adjacent to the clean room. Starting from June 2019, acquisition of the first data using the Staves
in the first available half-layer, IB-HL0 (innermost layer of the IB), began. Threshold, DAC tuning and noise occupancy scans are
periodically performed to monitor the performance of the detector. One example of threshold map obtained after the tuning
of the relevant DAC parameters and the effect of this tuning on the threshold distribution is visible in the right part
of Figure \ref{fig-2} for the IB-HL0. Still in the same Figure, the representation of the first reconstructed track of
a cosmic ray is shown. As can be seen from the detector sketch, the track reconstruction is possible thanks to the partial
overlap of adjacent Staves. 
It is also interesting to notice how the shape of the pixel cluster connected to the hit changes in the Staves 
sequentially crossed: in the first Stave (number 6), where the particle crossing is almost perpendicular, the cluster is small; 
in the second and third Staves (numbers 7 and 8), where the angle between the track and the chip plane decreases, the cluster 
becomes bigger and starts to resemble a piece of track.

From July 2019 the detector is kept powered and running; continuous monitoring is provided by three
shift crews alternating all along the 24 hours. Software development and hardware integration proceed
in parallel as well as preparation of all the infrastructures in the experimental area for the final installation.

\end{document}